\documentclass[12pt]{iopart}
\usepackage{iopams}
\expandafter\let\csname equation*\endcsname\relax
\expandafter\let\csname endequation*\endcsname\relax
\usepackage{graphicx}
\usepackage{amsmath}
\usepackage{amssymb}
\usepackage{amsfonts}
\usepackage{amsbsy}
\usepackage{color}

\renewcommand{\vec}[1]{\mbox{\boldmath$\mathrm{#1}$}}

\begin{document}

\title{Ultrafast transient dynamics in composite multiferroics}

\author{Chenglong Jia$^{1,2}$, Ning Zhang$^{1}$, Alexander Sukhov$^2$,  and Jamal Berakdar$^{2}$}

\address{$^1$Key Laboratory for Magnetism and Magnetic Materials of MOE, Lanzhou University, Lanzhou 730000, China}

\address{$^2$Institut f\"ur Physik, Martin-Luther Universit\"at Halle-Wittenberg, 06099 Halle, Germany}


\begin{abstract}
We investigate theoretically  the dynamic  multiferroic response of coupled ferroelectric/ferromagnetic  composites upon  excitation
 by a  photo-induced acoustic strain pulse. Two magnetoelectric mechanisms are considered: interface strain- and charge-mediated magnetoelectric couplings.
The former results in  demagnetization,  depolarization and repolarization within tens of picoseconds via respectively magnetostriction and piezoelectricity.
Charge magnetoelectric interaction  affects the ferroelectric/ferromagnetic feedback response leading to  magnetization recovery. Experimental realization based on time-resolved x-ray diffraction is suggested. The findings indicate the potential of composite multiferroics for
 photo-steered, high-speed, multi-state electronic devices.
\end{abstract}

\pacs{77.55.Nv, 61.05.cp, 78.47.J-, 77.22.Ej}

\vspace{2pc}
\noindent{\it Keywords}: muliferroic response, transient dynamics, demagnetization


\maketitle

\section*{Introduction}

\begin{center}
\begin{figure}[b]
\centering
\includegraphics[width=0.75 \textwidth]{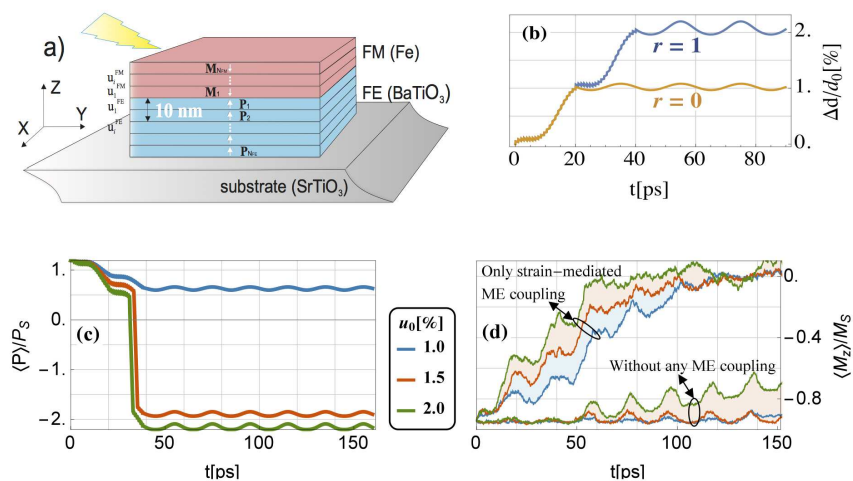}
\caption{(a) Schematics  of the proposed setup:
few ferromagnetic (FM) layers (e.g., Fe) coupled to a ferrelectric (FE) film (e.g., BaTiO$_3$,  or PbZr$_{1-x}$Ti$_x$O$_3$ deposited on a SrTiO$_3$ (STO)  substrate.
The structure
is irradiated with a laser pulse that induces a lateral acoustic  wave ($u_{i}(t)$ with laser tuneable amplitude $u_0$) triggering strain-driven multiferroic dynamics mappable by tracing the time evolution of  transient  FE polarization (c) (via trXRD), and FM magnetization (d) (via time resolved  magneto-optical Kerr effect). (b) Structure dynamics of the heterostructure with the reflectivity ($r$) from the substrate being $r=0$ or $r=1$. The general structure deformation is a super-position of these two cases.
}
\label{fig:System}
\end{figure}
\end{center}

Appropriately synthesized   ferroelectric (FE)  and ferromagnetic (FM)  multilayer or nano  structures may show
a multiferroic  (MF) (magnetic, electric, and/or elastic) response  which is indicative of an emergent coupling between the respective  order parameters
\cite{Multiferroics,Ramesh:2007fl,MTJ1, Composite-ME, Spaldin:2010tn,Velev:2011gu,Arrays-ME, MTJ2,Vaz:2012dp,Fusil:2014em,Liu:2014gy,Tokura:2014eg}. In addition to the fundamental questions regarding the origin of the underlying physics, this observation
holds the promise  of qualitatively new device concepts. Multiferroic memory devices \cite{ME-memory} with multi-state data storage and heterogeneous  read/write capability through the interfacial strain effects  \cite{Binek-07,Taniyama-09,Taniyama-11,Bibes-11,Berakdar-12}, the direct electric field effects\cite{Tsymbal-06,Mertig-10,apl13,chot13,Demkov-10,MeKl11,mAnst,Berakdar-10,Jia:2014in}, and exchange-bias \cite{Binek-05,Laukhin-06,Wu-10} are a few examples.
A key element thereby is the strength and symmetry of MF coupling and whether it is utilizable for swiftly
transferring/converting  FM into FE information.  Time-resolution, particularly  how fast such a conversion may take place and how to map it  in practice  are issues that
have not been addressed yet theoretically for MF composites, despite the intense research on MF materials. This work contributes to this
 aspect by  making a specific proposal for an experiment and provides theory and numerical simulations to unveil the time scale of mediating information (excitation) via MF coupling.
 Recently, first time-resolved  x-ray diffraction (trXRD) experiments were conducted to access the time-resolved FE response and lattice dynamics in single phase MF BiFeO$_3$ film \cite{schlom,schick}.
   Photo-induced stabilization and enhancement of FE polarization were observed for Ba$_{0.1}$Sr$_{0.9}$TiO$_3$/La$_{0.7}$Ca(Sr)$_{0.3}$MnO$_3$ \cite{sheu}. Our focus here is on  layered  FE/FM composites (cf.~Fig.\ref{fig:System}) whose magnetoelectric (ME) interaction may stem from the interfacial strain effects and/or spin rearrangement \cite{Composite-ME,Velev:2011gu,Vaz:2012dp,Jia:2014in}.

For isolated FM  systems, the ultrafast laser-induced magnetization dynamics, i.e. roughly speaking, a femtosecond demagnetization, a picosecond recovery, and a picosecond to nanosecond magnetization precession and relaxation are intensively studied with important   implications for photo-magnetic devices \cite{RMP-M}.
  For FE  nanostructured materials,   the ultrafast mechanical and electronic dynamics is
   well documented \cite{X-science,Bargheer-04}.  An optical pump excitation  pulse
  generates a propagating mechanical stress, which results in picosecond polarization dynamics \cite{Bargheer-06,Bargheer-07,Woerner:2009fr} that can be probed
   experimentally via trXRD. In a composite \textcolor{black}{MFs}, is yet to be clarified how upon such a pump pulse  the coupled time-resolved MF dynamics is manifested, an issue addressed here.
   %
 %
%
%

 \section*{Generalities and proposed setup}

Experimentally, ferroelectric/ferromagnetic (FE/FM) multiferroic heterostructures were successfully realized and characterized \cite{Multiferroics,Ramesh:2007fl,MTJ1, Composite-ME,Vaz:2012dp,Fusil:2014em,Liu:2014gy}. In principle, strain and charge co-mediated magnetoelectric coupling are expected in composite MFs \cite{Nan:2014ck}.
The direct charge-mediated magnetoelectric interaction is generally due to the induced changes in the magnetic states by the electrostatic screening effect \cite{Vaz:2012dp,Jia:2014in}, it is however strong and plays a dominant role in some FE/FM-metal systems  \cite{Mardana:2011cj,Jedrecy:2013bc,Zhang:2015hs,Jia:2015iza}. Whereas, the piezoelectric strain is found to give rise to an electrically tunable  uniaxial magnetic anisotropy \cite{Vaz:2012dp,Berakdar-12,Nan:2014ck}. 
 To unveil the transient  dynamics  we propose Fig.(\ref{fig:System}) to employ
 photo-induced transient strain and trXRD and monitor  the effects of interface strain- and charge-mediated ME couplings. The strain can be chosen lateral, as in our case, or as having
  in-plane components by an additional appropriate grating atop the FM film.  For  strain-induced
 magnetization dynamics in conventional isolated FM we refer to Ref. \cite{strain_ferromagnets} and references therein.
 Calculations show, for a multiferroic composite chain electrically-induced magnetization reversal is not achievable for large/thick FM subsystem \cite{Berakdar-11,Berakdar-12}. This is because of the interface-limited
  nature of  MF coupling \cite{Vaz:2012dp,Jia:2014in}. Therefore, it is advantageous  to choose a system
    consisting of a thin FM layer  (such as Fe with thickness $d_{\text{FM}} =10$ nm) and a thicker FE layer (e.g. PbZr$_{1-x}$Ti$_x$O$_3$ (PZT) or BaTiO$_{3}$ \textcolor{black}{(BTO)} with thickness $d_{\text{FE}} =100$ nm)  grown epitaxially on a substrate \textcolor{black}{SrTiO$_3$} (STO) [Fig.\ref{fig:System}(a)].  Fe[110] can be  caped (with
    thin transplant Au-layer to prevent oxidation) and rotated to align  parallel to BTO[100], resulting in  in-plane misfit strains $u^{\text{Fe}}_{\|} = 1.39 \%$ and $u^{\text{BTO}}_{\|} = -0.139 \%$ \cite{footnote-strain}.
    The normal strains are determined by the Poisson ratio, $n_{\text{Fe}} = u_{\|}^{\text{Fe}}/ u_{z}^{\text{Fe}}$ and  $n_{\text{BTO}} = u_{\|}^{\text{BTO}}/ u_{z}^{\text{BTO}}$ with $n_{\text{Fe}} = 0.26$ \cite{Book-Ibach} and $n_{\text{BTO}} = 0.65$ \cite{Neaton-MRS-2002}, respectively.  
In the following we focus on the particular
situation where the spontaneous FE polarization is directed perpendicular to the substrate plane (hereafter referred to as the $\vec{e}_{z}$-direction).

From a computational point of view, the coarse-graining procedure with cell size $a=5$ nm is conveniently used to obtain the macroscopic quantities  of polarization $\vec{P}_{i}$ (with $i = 1, ..., N_{\text{FE}}$ and $N_{\text{FE}} = d_{\text{FE}}/a$) and magnetization $\vec{M}_{j}$ (with $j = 1, ..., N_{\text{FM}}$ and $N_{\text{FM}} = d_{\text{FM}}/a$) \cite{Horley:2012ky}. The change in polarization $\vec{P}_{i}$ to a first-order approximation can then be understood in terms of  piezoelectricity \cite{Piezoe-1},
\begin{equation}
\Delta P_{iz} = \sum_{\xi} c_{3 \xi} u_{i\xi}^{\text{FE}} +\epsilon_{0} \chi^{e} E_{i},
\label{eq:Piezoelectricity}
\end{equation}
where $c_{3\xi}$ ($\xi =1, 2, 3$ ) is the improper piezoelectric tensors \cite{Piezoe-2}, $\epsilon_{0}$ is the free space permittivity, and $\chi^{e}$ is the electric susceptibility. The effective electric field $E_{i}$ derives as  $E_{i} = \delta F_{\text{FE}}/\delta \vec{P}_i $, where $F_{\text{FE}}$ is being the coarse-grained FE free energy \cite{Berakdar-12}, $F_{\text{FE}} = F_{\text{EE}} + F_{\text{DDI}}$.
%
%
The elastic Gibbs function $F_{\text{EE}}$ corresponding to the tetragonal phase of BaTiO$_{3}$ reads  \cite{Berakdar-12,Sukhov:2013cea}
\begin{eqnarray}
&& F_{\text{EE}} = - \frac{\alpha}{2} \sum_{i} \vec P_i^2 + \frac{\beta}{4} \sum_{i}  \vec P_i^4 + \kappa \sum_{i} (\vec P_i -\vec P_{i-1})^2 \\
&&   + \mathrm{{\frac{1}{\epsilon_0}}} \sum_i \left[c_{31} (u^{\text{FE}}_{ix} +u^{\text{FE}}_{iy}) + c_{33} u^{\text{FE}}_{iz} \right] P_{iz} +  \mathrm{{\sum_i}} \frac{1}{2} C_{11}^{FE} \vec{u}^{\text{FE}}_{i} \cdot \vec{u}^{\text{FE}}_{i}. \nonumber
\end{eqnarray}
Here we accounted for the symmetry $c_{31} = c_{32}$.  The stiffness coefficient of the FE part is $C_{11}^{\text{FE}}$. $F_{\text{DDI}}$ is the long range FE dipole-dipole interaction which has the usual form
\begin{equation}
F_{\text{DDI}} = \frac{1}{4\pi \epsilon_{\text {\text{FE}}} \epsilon_0}\sum_{i\neq k} \left[ \frac{\vec{P}_i \cdot \vec{P}_k - 3(\vec{P}_i \cdot \vec{e}_{ik})(\vec{e}_{ik}\cdot \vec{P}_k)}{r_{ik}^3} \right].
\end{equation}
where $\epsilon_{\text{FE}}$ is the FE permittivity,
 $r_{ik}$ is the distance between  $\vec{P}_i$ and $\vec{P}_k$, and $\vec{e}_{ik}$ is the unit vector joining the two dipoles.

Analogously, for  the ferromagnetic energy density the relation  applies $F_{\text{FM}} = F_{\text{XC}} + F_{\text{MMI}}$.
%
%
$F_{\text{XC}}$ consists of the nearest-neighbor exchange \textcolor{black}{interaction} (A-term) between
 $\vec{M}_{j}$ \textcolor{black}{and $\vec{M}_{j+1}$}, the uniaxial magneto-crystalline anisotropy contributions ($K_{1}$-term), and the (magneto-) elastic energies,
\begin{eqnarray}
F_{\text{XC}} &=& - \frac{A}{a^2M_S^2} \sum_{j} \vec{M}_j \cdot \vec{M}_{j+1} - \frac{K_1}{M_s^2} \sum_{j} \vec{M}_{jz}^2  \\
&+& \mathrm{{\frac{B_1}{M^{2}_S}}} \sum_{j,\xi} u^{\text{FM}}_{j\xi} (M_{j\xi}^{2}- 1/3) +\frac{1}{2} \sum_{j}C_{11}^{\text{FM}} \vec{u}^{\text{FM}}_{j} \cdot \vec{u}^{\text{FM}}_{j}. \nonumber
\label{eq:XC}
\end{eqnarray}
$M_s$ is the saturation magnetization. The anisotropy $K_{1}$ depends on the FM film thickness $d_{\text{FM}}$, $K_{1} = \left( {K_{s}}/{d_{\text{FM}}} - {\mu_{0} M_{s}^{2}}/{2} \right)$, where $K_{s}$ describes the surface anisotropy contributions that are significant for ultra-thin film tending to align the magnetization normal to the surface, whereas, $\mu_{0}M_{s}^{2}/{2}$ denotes the demagnetization field that is equivalent to an easy in-plane contribution. $B_{1}$ and $C_{11}^{\text{FM}}$ \textcolor{black}{respectively} denote  the magneto elastic constants and elastic moduli of the FM layer.  The magnetic dipole-dipole interaction \textcolor{black}{$F_{\text{MMI}}$} is
\begin{equation}
F_{\text{MMI}} = \frac{\mu_0}{4\pi}\sum_{j\neq l} \left[ \frac{\vec{M}_j \cdot \vec{M}_l - 3(\vec{M}_j \cdot \vec{e}_{jl})(\vec{e}_{jl}\cdot \vec{M}_l)}{r_{jl}^3} \right],
\end{equation}
where $\mu_0$ is the magnetic permeability constant.

We are targeting exclusively  ps-ns time scales, i.e. beyond
the range for laser-induced fs-demagnetization \cite{RMP-M} is not discussed here. The dynamics of orbital degrees of freedom is therefore not explicitly taken into account (i.e., they are assumed to have relaxed to the dynamics considered here). The material parameters for the FE subsystem are chosen as $\alpha = 2.77 \times 10^7$ Vm/C \cite{Parameters-FE}, $\beta = 1.70 \times 10^8$ Vm$^5$/C$^3$ \cite{Parameters-FE}, $\kappa = 1.0 \times 10^8$ Vm/C \cite{Berakdar-10},  $P_d = 0.499$ C/m$^2$ \cite{Berakdar-12}, and $C_{11}^{FE} = 1.78 \times 10^{11}$ N/m$^2$ \cite{Book-FE}. The improper piezoelectric constants are set as those of BTO \cite{Piezoe-2}: $c_{31} = 0.3$ $C/m^2$, $c_{33} = 6.7$ C/m$^2$, and the Poisson ratio $n=0.64$ \cite{nPoisson}.  Further material parameters concerning the  FM layer are  iron, i.e., $\lambda = 2.07 \times 10^{-5}$ along Fe [100], $B_1= -2.95 \times 10^6$ N/m$^2$, $C_{11}^{FM} = 2.41 \times 10^{11}$ N/m$^2$\cite{Book-FM1}, $A = 2.1 \times 10^{-11}$ J/m \cite{Book-FM2}, $K_1 = 4.8 \times 10^4$ J/m$^3$ \cite{Book-FM2}, $M_s = 1.71 \times 10^6$ A/m \cite{Book-FM2}. We assume that none of these parameters changes during fast dynamics of interest here.

From the symmetry point of view, the space-inversion symmetry and the time-reversal symmetry are intrinsically broken at the FE/FM interface \cite{MF-Scott,MF-Spaldin}, the multiferroic coupling is thus restricted to the region in the vicinity of the interface,
\begin{equation}
F_{\text{ME}} = \gamma \vec{P}_{1} \cdot \vec{M}_{1} - \frac{3}{2} \lambda \sigma \cos^2 \phi.
\label{ME-coupling}
\end{equation}
%
%
The first $\gamma$-term takes its origin from a magnon-driven, direct ME interaction in the vicinity of the FE/FM interface acting within the spin-diffusion length on the order of  nanometers \cite{Jia:2014in,Jia:2015iza}. $\gamma$ is the coupling strength in unit of s/F. Given that the spin-diffusion length is around $8.5$ nm in Fe \cite{Bass:2007bw},  the linear direct ME coupling is assumed to only involve in the interfacial nearest neighbor cells, the FE polarization $\vec{P}_{1}$ and FM magnetization $\vec{M}_{1}$, respectively. The second $\lambda$-term involves
 piezoelectricity and the magnetostriction at the interface,   associated with an additional uniaxial anisotropy energy for the FM layer \cite{Binek-07,Berakdar-12,Book-FM1}.   $\lambda$ is the average magnetostriction coefficient, and $\phi$ is the angle between the magnetization $\vec{M}$ and the direction of the stress $\vec{\sigma}$ across the interface,
%
$\vec{\sigma} = -C_{11}^{\text{FE}} \vec{u}^{\text{FE}}_{1}+ C_{11}^{\text{FM}}\vec{u}^{\text{FM}}_{1}$.
%
The in-plane \emph{static} stress at the interface is assumed to be balanced due to the lattice deformation. Taking that $\lambda >0$ in our case, a negative film stress ($\sigma < 0$) on FM layer favors $\phi=\pi/2$ which means an in-plane magnetization while $\sigma >0$ favors an out-of-plane magnetization with $\phi=0$ .  It should be noted that we focus on thin FM films, the indirect high-order ME coupling, such as a spin-motive force resulting from the non-equilibrium magnetic domain wall dynamics \cite{Barnes:2007ko}, is disregarded here.  Other cases of strains can be treated  similarly.

As demonstrated by  trXRD experiments \cite{Bargheer-04,Bargheer-06,Bargheer-07,Bargheer-09},  except for a
static strain due to the lattice mismatch,  electronic excitation by an ultrafast pump pulse  generates \emph{dynamic} transient strain propagating through the FE/FM films, directly affecting the dynamics of the magnetization and polarization. Here we presume that an applied optical pulse is exclusively absorbed in  Fe layer,  changing so the electronic configuration of the absorbing material and generating a transient stress that results in a displacive excitation of phonons in the Fe/BTO systems through the electron-phonon coupling.
The expansion front with an amplitude $u_{f}(0,0)$ starts at time $t=0$ from the top air-Fe interface, enters into the Fe and BTO layers with the respective sound velocity ($v_{\text{Fe}} = 5130$ m/s and $v_{\text{BTO}} = 5437$ m/s, respectively)  and arrives at the surface of the substrate after $\Delta t = d_{\text{FM}}/v_{\text{Fe}} + d_{\text{FE}}/v_{\text{BTO}} \approx 20$ ps.  At the BTO/substrate interface, the strain front is reflected 
from the substrate surface and \textcolor{black}{backs} into the Fe/BTO heterostructure,  encountering  the incoming strain wave,  and then launching a coherent acoustic standing wave with the wave vector $k = 1/(d_{\text{FM}} + d_{\text{FE}})$. Such a coherence lattice motion is manifested  in  fast oscillation (with system-size-determined period $T= \Delta t $) of FE polarization and FM magnetization due to  piezoelectricity and  magnetostriction respectively, as shown in Fig. \ref{fig:System} and Fig. \ref{fig:MP-charge}. Furthermore, the polarization/magnetization dynamics are coupled to each other at the interface due to the ME coupling, which  gives rise to marked changes in FE/FM response since the surface contribution to the free energy plays an important role for nanostructures.

The time evolution of the strain wave depends significantly on the  pump fluences \cite{Bargheer-06}.  Without loss  of  generality
two  limiting cases are to be considered:\\
\emph{(i) Strong excitation.} For large  pump fluences, the heterostructures may suddenly deform within femtoseconds. For ps dynamics, the
 strain front amplitude acts promptly as  $u_{f}(z,t) = -u_{0}  \sin(2\pi kz - 2 \pi \omega t +\pi/2) $ (i.e.~,$-u_{0}$ for $t=0$) with the frequency $\omega = 1/T$.
  The strain standing wave reads
%
$u_{s}(z,t) = -2u_{0} \left[ 1 - \cos(2\pi kz) \sin (2\pi \omega t) \right]$ 
after $t = 2T = 40 ~ps$.  \\
\emph{(ii) Weak excitation.} For  moderate pump fluences  the expansion front $u_{f}(z,t) = u_{0} \sin(2\pi kz -2\pi \omega t)$ travels within  FE/FM heterostructures leading to the standing wave
%
 $u_{s}(z,t) = 2u_{0}  \sin(2\pi kz) \cos (2\pi \omega t).$
%
The general pump case is  a super-position of  these two cases.

\section*{Numerical results and analysis}
The multiferroic dynamics is studied by  kinetic Monte Carlo simulations \cite{Book-MC,Berakdar-12} with open boundary condition at room temperature ($300$ K) for tetragonal BTO phase. \textcolor{black}{The kinetic Monte Carlo method is advantageous in that  it is computationally more tractable  than a direct solution of the coupled Landau-Lifshitz-Gilbert/Landau-Khalatnikov equations that we examined earlier \cite{Berakdar-10}.} The magnetic moments $\vec{M}_j$ are understood as three-dimensional unit vectors, which are updated coherently, \emph{i.e.}, at each trial step \textcolor{black}{the} direction of new $\vec{M}_j$ is limited within a cone around the initial spin direction \cite{Hinzke-98}.  The maximum angle $\theta_{max}$ of the cone is determined by means of a feedback algorithm so that the number of accepted spin modifications is just half the total number of equilibrium configurations at a given temperature before the x-ray diffraction \cite{SGL-93}. In experiments with BTO, the FE dipoles in the tetragonal phase are along the [001] direction and  are thus assumed to be bi-directional vectors.  The remanent polarization is $\vec{P}_s$ and the field-induced deviation is $\Delta P_i$ and is given by Eq. (\ref{eq:Piezoelectricity}) \cite{FE-MC1,FE-MC2}.  During the simulations, the multiferroic equilibrium at $300$ K is at first established with $\Delta P_i =0$, $\theta_{max}$ is determined. Then a transmit strain along the chain is turned on at $t=0$ and it propagates through the multiferrroic chain. The strain wave and all induced FE dipole moments are updated with the time step $\tau_{0} = 0.1 ~ ps$, which is also taken as the \emph{time unit} of Monte-Carlo algorithm \cite{Book-MC}. To reduce possible random errors, the data are collected and averaged for 4000 independent runs.

As discussed after Eq. (\ref{ME-coupling}), the MF dynamics is in general strain and charge mediated.
 However, for the type of excitations considered here strain is a key factor.
 For an insight into various mechanisms, at first only  strain-mediated ME interaction is considered. Fig. \ref{fig:System} shows transient changes in the averaged polarization and magnetization  for a strong  pump pulse with 100\% reflectivity from the substrate.  FE polarizations are strongly suppressed by the photo-induced stress (Fig. \ref{fig:System} (c)), with/without the strain-mediated ME couplings. There are two distinguishable change steps (at $t=20$ ps and $t=40$ ps, respectively) corresponding to $\pi$-phase shifts between the incoming expansion wave and the reflected front before forming a standing strain wave. As the peak strain exceeds $1.4\%$ a critical point is arrived. The negative piezoelectric contribution $\Delta P_i$ exceeds the permanent dipole $\vec P_{s}$,
 a full FE polarization reversal is then induced by the strong piezoelectricity within $40 ~ps$. Experimentally such a strain-induced ultrafast characterization of polarization dynamics \textcolor{black}{has} indeed been observed in a PbZr$_{0.2}$Ti$_{0.8}$O$_3$/SrRuO$_3$ superlattice \cite{Bargheer-06,Bargheer-07,Bargheer-09}, where the FE dynamics was traced back to the anharmonic coupling of the tetragonal distortion and the ferroelectric soft phonon mode in PZT.  For the FM subsystem in general,  an
 optical excitation generates a time-dependent magneto-elastic anisotropy making the normal $z$ axis  magnetically harder with the increase of the amplitude of strain wave $u_0$ (cf. Eq. (4)). This leads to a fast-oscillation but a relatively weak and slow reduction of the normal magnetization $\langle M_z \rangle$ (c.f. Fig. \ref{fig:System} (d)).
Upon accounting for the uniaxial interfacial magnetic anisotropy ($\lambda$-term in Eq. (\ref{ME-coupling}))  that stems
 from the stress $\sigma$ across the FE/FM interface, we find a magnetic transition from ferromagnetic to paramagnetic state (i.e., $\langle M_{z} \rangle \rightarrow 0$) in  $100 ~ps$ range (Fig. \ref{fig:System} (d)). We recall that we adopt a phenomenology based on coarse-grained order parameters that formally result from an averaging over microscopic quantities over an $5\times 5 \times 5 ~ nm^3$ cell.
As mentioned above the time scale and the origin of the demagnetization processes in our system \textcolor{black}{are} quite different from the conventional  fs to ps laser-induced demagnetization dynamics  \cite{RMP-M,L-LLG}.
 Furthermore, the lattice deformation $u_i < 0$ results in a surface tensile rather than compressive strain to Fe along the z-axis. So, such a collapse of magnetic order does not  correspond to the case of \emph{"Iron under pressure"} \cite{Fe-pressure-1, Fe-pressure-2,Fe-pressure-3}, where magnetic transition is simultaneously accompanied by a high \emph{pressure} bcc to hcp structure transition. Here the magnetic collapse phenomenon is attributable to the \emph{extraordinarily} hardening  of the magnetic uniaxial $\vec \sigma$-axis.
As strain wave propagates through the heterostructures, the interface stress $\vec \sigma$ acting on Fe is rapidly oscillating and reaches a giant value, for instance $\sigma = -16.5 $ GPa with $u_0 =1\% (2\%)$ at $t =55(15)$ ps, which makes the stress axis  extremely hard,
 altering substantially the ferromagnetic order along the magnetocrystalline axis within tens of picoseconds. The in-plane magnetization is then favorable.
  Considering  the SO(2) rotational symmetry of the magnetoelastic anisotropy \textcolor{black}{($B_1$-term in Eq. (4))} about the $\vec{e}_{z}$-axis, however, there is no preferred easy axis in the normal plane to the direction of surface stress, resulting in a rotational in-plane anisotropy. The numerical calculations confirm that the in-plane averaged magnetization $\langle M_x \rangle$ and $\langle M_y \rangle $ present a noise-like dynamic behavior, we thus have an ultrafast interface strain-driven demagnetization in a multiferroic FE/FM heterostructure.

\begin{figure}[t]
\centering
\includegraphics[width=0.75 \textwidth]{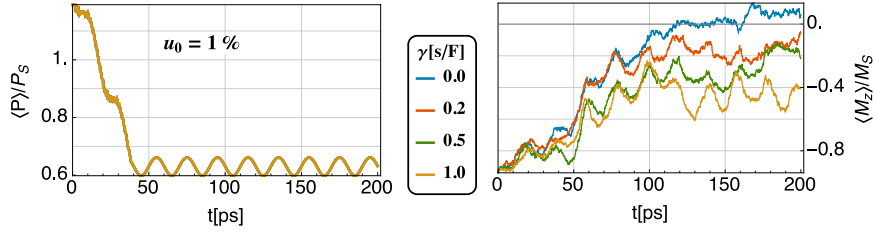}
\caption{Polarization and magnetization feedback dynamics induced by  interplay of
interface strain and charge co-mediated magnetoelectric couplings  (with varying strength $\gamma$) for  strong excitation with $u_{0} =1\%$. Comparing with the amplitude of magnetization in Fig.\ref{fig:System} (d) after 100 ps, an obvious magnetization recovery is induced by the charge-mediated ME interaction.}
\label{fig:MP-charge}
\end{figure}

For a more complete scenario of ultrafast MF dynamics, another interfacial ME coupling induced by the spin-polarized screening charges should be considered \cite{Jia:2014in}.   Different from  strain-mediated MF interaction, the charge-mediated, magnon-driven ME  effect couples directly the dynamics of FE polarization and FM magnetization and favors antiparallel alignment perpendicular to the interface. Taking into account that the strength of direct ME interaction is around 1 s/F in metallic FM film \cite{Jedrecy:2013bc,Jia:2014in,Jia:2015iza}, such electrically controllable effective magnetic field generated by the non-vanishing normal FE polarization gives rise to an induced magnetization along $\left<P_{\mathrm{z}}\right>$ from the paramagnetic state, as evidenced in Fig. \ref{fig:MP-charge}. On the other hand, with  strain amplitude $u_0 =1\%$, the coupling strength $\gamma = 1$  s/F is insufficient to produce considerable feedback changes in FE polarization. The FE/FM feedback is only pronounced, for a large coupling $\gamma$. The pre-contact FE cell diminishes in magnitude and then flips its direction for $\gamma > 14$  s/F, causing the emergence of the re-polarized FE. Simultaneously, due to the interplay of the interface strain and charge-mediated magnetic anisotropy, the FM part recovers remnant magnetism, which favors the opposite orientation of the polarization as expected.

\begin{figure}[ht]
\centering
\includegraphics[width=0.75 \textwidth]{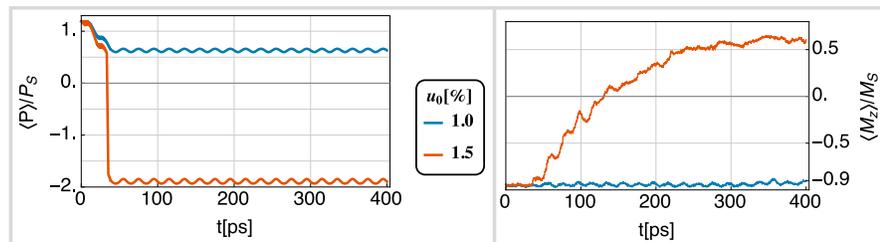}
\caption{Multiferroic responses with only the charge-mediated interface magnetoelectric coupling  ($\gamma =1.0$ s/F and {$\lambda=0$}). The FE repolarization (left) gives rise to a magnetization reversal (right).}
\label{fig:Charge-only}
\end{figure}

Given a large mismatch in stiffness coefficients between FE/FM films \cite{mAnst} and/or appropriately fabricating FM/FE crystal orientation \cite{Zhang:2015hs}, the magnetostrictive effects can be minimized, the direct electric-field effects would then dominate the multiferroic dynamics. In Fig. \ref{fig:Charge-only} the multiferroic responses, driven by charge-mediated ME interaction \emph{only}, are demonstrated. As one can see, the propagating mechanical stress with the strain-front $1.5\%$ reverse the FE polarization and consequently the FM magnetization due to the requirement of the antiparallel configuration between the polarization and magnetization by the direct charge-mediated ME coupling.
%

\begin{figure}[ht]
\centering
\includegraphics[width=0.85 \textwidth]{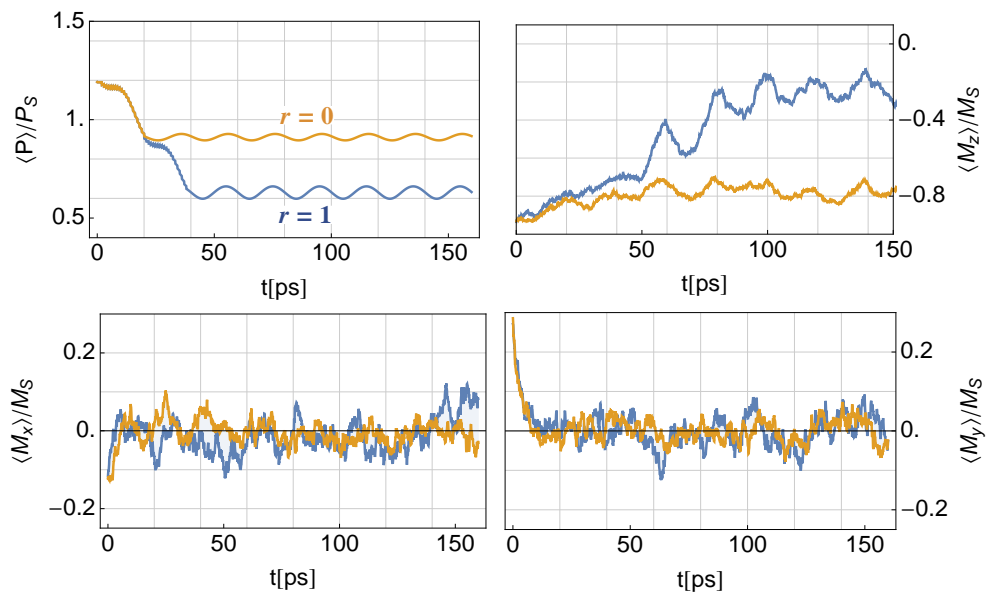}
\caption{Multiferroic dynamics for  weak excitation  ($u_{0} =1\%$) with different reflectivity ($r =0 ~\text{or} ~ 1$) from the substrate. A giant interface charge-mediated magnetoelectric coupling $\gamma = 25$ s/F is assumed in the theoretical simulation (not appropriate for FE/FM heterostructures in experiments).}
\label{fig:MP-weak}
\end{figure}

To explore the FE/FM feedback response in the case of \emph{weak excitation},  a very large interface charge-mediated magnetoelectric coupling
 $\gamma = 25$ s/F  is assumed in the theoretical simulation. The system exhibits demagnetization due to the dynamic strain effect  as well (cf. Fig. \ref{fig:MP-weak}), but FE re-polarization vanishes even with such an unrealistic (giant) interface charge-mediated ME coupling, though the FE/FM feedback are still present and induces a quasi-square-pulse FE dynamic behavior.

\section*{Conclusions}

We proposed and  theoretically realized a scheme for studying ultrafast dynamics in a composite MF heterostructure related to Fe/BaTiO$_3$.
Based on the piezoelectricity and the magnetostriction resulting from a coherent lattice motion in the FE and the FM launched by a pump laser pulse, the critcial amplitude of the strain front $1.4\%$ gives rise to a complete switching of the bipolar FE polarization, while at the same time for zero ME coupling of any type it has low impact on the FM order (Fig. \ref{fig:System} (c), (d)). Only in the presence of the strain-mediated ME coupling (second term of Eq. (\ref{ME-coupling})) the total out-of-plane magnetization becomes supressed (Fig. \ref{fig:System}, (d)). The effect of the charge-mediated coupling (first term of Eq. (\ref{ME-coupling})) is opposite, i.e. it results in a partial recovery of the total $M_{\mathrm{}z}$-component on the time scale of about 200 ps (Fig. \ref{fig:MP-charge}).
It should be noted that the reflectivity of propagating strain wave from the substrate is not necessary for the ultrafast multiferroic dynamics but it indeed enhances the studied effects (cf. Fig. \ref{fig:MP-weak}).
In addition, it is numerically evident that such ultrafast magnetoelectric dynamics are general in FE/FM heterostructures containing strong piezoelectric ferroelectric subsystem, such as BaTiO$_{3}$ and PbTiO$_{3}$ \cite{Piezoe-2}.  The results indicate the potential of MF composite  for photo-operated high-speed devices.
%

\ack{Discussions with M. Alexe, T. Els\"asser, and D. Hesse  are gratefully acknowledged.
This work was supported by  the National Basic Research Program of China (No. 2012CB933101), the National Natural Science Foundation of China (No. 11474138), the German Research Foundation (No. SFB 762), the Program for Changjiang Scholars and Innovative Research Team in University (No. IRT1251), and the Fundamental Research Funds for the Central Universities.}


\newpage

%

\begin{thebibliography}{99}
\bibitem{footnote-strain} The in-plane misfit strains are estimated by, $u^{Fe}_{\|} = -\big( a_{\|} - a^{Fe}_{0}\big)/a_{\|} = 1.39 \%$ and $u^{BTO}_{\|} = -\big(a_{\|} - a^{BTO}_{0}\big) /a_{\|} = -0.139 \%$, where $a^{Fe}_{0}$ ($0.4053$ nm along Fe[110]) and $a^{BTO}_{0}$ ($0.3992$ nm along BTO[100]) are free lattice constant of Fe and BTO, respectively. $a_{\|} \approx \frac{a^{Fe}_{0} d_{Fe} + a^{BTO}_{0}d_{BTO}}{d_{Fe}+d_{BTO}} = 0.3998 ~nm$  is the in-plane lattice constant of epi-bilayer,
\end{thebibliography}
\end{document}